\documentclass{emulateapj}

\begin{document}
\title{DEEP CHANDRA OBSERVATION OF THE PULSAR WIND NEBULA 
POWERED BY THE PULSAR J1846--0258 IN THE SUPERNOVA REMNANT KES 75}

\author{C.-Y. Ng\altaffilmark{1},
P. O. Slane\altaffilmark{2}, B. M. Gaensler\altaffilmark{1,\dag},
J. P. Hughes\altaffilmark{3}}
\altaffiltext{1}{Institute of Astronomy, School of Physics, The University of Sydney, NSW 2006, Australia.}
\altaffiltext{2}{Harvard-Smithsonian Center for Astrophysics, 60 Garden Street, Cambridge, MA 02138, USA.}
\altaffiltext{3}{Department of Physics and Astronomy, Rutgers University, 136
Frelinghuysen Road, Piscataway, NJ 08854, USA.}
\altaffiltext{\dag}{ARC Federation Fellow}
\email{ncy@physics.usyd.edu.au}

\begin{abstract}
We present the results of detailed spatial and spectral analysis of the
pulsar wind nebula (PWN) in supernova remnant Kes 75 (G29.7--0.3) using
a deep exposure with Chandra X-ray observatory. The PWN shows a complex
morphology with clear axisymmetric structure. We identified a one-sided
jet and two bright clumps aligned with the overall nebular elongation,
and an arc-like feature perpendicular to the jet direction. We interpret
the arc as an equatorial torus or wisp and the clumps could be shock
interaction between the jets and the surrounding medium. Spatial modeling
of the features with a torus and jet model indicates a position angle
$207\arcdeg\pm8 \arcdeg$ for the PWN symmetry axis. The lack of any observable
counter jet implies a flow velocity larger than 0.4$c$. Comparing to an
archival observation 6 years earlier, some small-scale features in the
PWN demonstrate strong variability: the flux of the inner jet doubles,
and the peak of the northern clump broadens and shifts 2\arcsec\ outward.
As recently reported from RXTE observations, magnetar-like bursts of the
central pulsar occurred coincidentally 7 days before the Chandra exposure.
Accompanied with the bursts is a temporary increase in the pulsar flux:
during the Chandra observation, the pulsar luminosity is 6 times larger
than in quiescent state, and shows substantial spectral softening from
$\Gamma=1.1$ to 1.9 with an emerging thermal component.
\end{abstract}


\section{INTRODUCTION}
\object{Kes 75} (G29.7--0.3) \citep{kes68} is a composite supernova
remnant (SNR) in our Galaxy that harbors one of the youngest known
neutron stars.  At the center of the 3\arcmin\ diameter SNR shell,
there is a pulsar wind nebula (PWN) powered by the source \object{PSR
J1846--0258}. In radio frequencies, the VLA and the
Berkeley-Illinois-Maryland Association (BIMA) array images show a
diffuse SNR shell with a flat spectrum core \citep{hcg03,bg05}.
X-ray observations with ASCA revealed the thermal and non-thermal
spectral components of Kes~75 \citep{bh96}. A 37\,ks Chandra exposure
resolved the shell structure and discovered a neutron star embedded
in an axisymmetric PWN with complex morphology \citep{hcg03}. The
observation also indicated a remarkable coincidence between the X-ray
and radio emission. Comparing Spitzer infrared images to deep Chandra
observations, \citet{mor07} suggested a Wolf-Rayet progenitor for the
remnant. At higher energy, the PWN has been detected in soft $\gamma$-rays
by INTEGRAL \citep{mcb08} and in the TeV band by HESS \citep{dja08}.
The first distance estimate to the source placed it at 19\,kpc based on
neutral hydrogen absorption measurements \citep{bh84}, implying very high
efficiency with which the pulsar converts its spin-down power into X-ray
emission \citep{hcg03}. Recently \citet{lt08} deduced a new distance
estimate of 5.1-7.5\,kpc from H{\sc I} and $^{13}$CO maps, providing a
more reasonable pulsar luminosity and remnant size.

The pulsar J1846--0258 was discovered in the X-ray band with timing
observations using the Rossi X-ray Timing Explorer (RXTE) \citep{got00},
but remains undetected at radio frequencies \citep{arc08}. The spin period
($P=324$\,ms) and spin down rate ($\dot{P}=7.1\times10^{-12}$\,s\,s$^{-1}$)
imply a large spin-down luminosity of $\dot{E}\equiv 3.9\times10^{46}\dot{P}
/P^3$\,ergs\,s$^{-1}=8.1\times10^{36}$\,ergs\,s$^{-1}$, a high surface
magnetic strength $B\equiv 3.2\times10^{19}(P/\dot{P})^{1/2}$\,G$=4.9
\times10^{13}$\,G, and a small characteristic age $\tau_c\equiv P/2
\dot{P}=720$\,yr. The large inferred surface magnetic field is near the
values found for pulsars whose emission is magnetically dominated, i.e.\
the magnetars \citep{wt06}. Subsequent RXTE monitoring observations
indicated a braking index of $n=2.65\pm0.01$ \citep{liv06}, implying a
spin-down age $\tau\equiv P/(n-1)\dot{P}=884$\,yr, which makes it one of
the youngest known neutron star in our Galaxy \citep[see,][]{bb08}.
\citet{gav08} recently reported 5 short X-ray bursts from the pulsar in
2006, along with temporary changes in the spectral and timing noise
properties, suggesting magnetar-like nature of this object. Coincidentally,
the deep Chandra observation presented in this study was taken only 7
days after these unprecedented bursts. With the same dataset, \citet{ks08}
independently reported the increase in the pulsar flux and the softening
in its spectrum at this time, as a result of the bursts. We note that the
above studies focused on the central pulsar and left unresolved the question
of variability in the PWN.

In this paper, we concentrate on the spatial and spectral properties of the
PWN in Kes 75 using deep Chandra observations. Specifically, we investigate
the variability in the system by comparing exposures from different epochs.
This analysis provides a unique opportunity to study the evolution of a young
PWN and its interaction with the environment. The observations and data reduction
are described in \S\ref{s2}. In \S\ref{s3}, we present the spatial analysis
of the overall PWN structure. The variabilities of the pulsar and the PWN are
studied in \S\ref{s4}. The spectral analysis is presented in \S\ref{s5} and in
\S\ref{s6} we discuss the interpretations of all the above results.

\section{OBSERVATIONS AND DATA REDUCTION}
\label{s2}
Deep Chandra observations of Kes 75 were carried out on Jun 5-12, 2006 in four
exposures (Observation IDs [ObsIDs] \dataset [ADS/Sa.CXO#obs/06686] {6686}, \dataset
[ADS/Sa.CXO#obs/07337] {7337}, \dataset [ADS/Sa.CXO#obs/07338] {7338} \& \dataset
[ADS/Sa.CXO#obs/07339] {7339}) which began 7 days after the first four bursts
that were reported by \citet{gav08}, and 50 days before the fifth one. The
data were taken with the ACIS-S array\footnote{The ACIS-S detector has pixel
size of 0.492\arcsec\ and a practical detection range of 0.3-10\,keV in energy,
with spectral resolution from 100 to 250\,eV across the energy range (see
\url{http://cxc.harvard.edu/proposer/POG/html/chap6.html}).} operating in very
faint (VF) timed exposure imaging mode using a custom CCD subarray of 600 rows
(frame time=1.8s) in order to reduce the photon pile-up. The total exposure was
155\,ks and the pulsar was positioned near the aim point on the S3 chip, which
was the only CCD active during the observation. For comparison, we have also
reprocessed the archival 37\,ks exposure (\dataset [ADS/Sa.CXO#obs/0748]
{ObsID 748}) observed on Oct 15, 2000 \citep{hcg03}, which was carried out in
ACIS-S faint TE mode with the full CCD frame (frame time = 3.2s). All data reduction
was performed using CIAO 4.0 and CALDB 3.4.2. We started with the level=1 event
files, turned off the pixel randomization, and applied the time-dependent gain
calibration and charge transfer inefficiency (CTI) correction. We did not apply
the background cleaning algorithm for the VF observations, which uses the outer
16 pixels of the $5\times5$ event island to improve discrimination between good
events and likely cosmic rays, since this rejects too many real events from the
bright source. Examination of the background light curves showed no strong flares
during all the observations. Hence, all data are included in the analysis. For
imaging analysis, the four individual exposures of the 2006 observation were
co-added. We first determined the pulsar position with the task {\tt wavdetect},
then aligned the images accordingly. Assuming the roll angle in the header is correct,
we applied only linear shifts for the alignment. Since the amount of shift required
is at most 0.5 pixels (i.e.\ $\sim0.2\arcsec$), which is much smaller than the
features we are interested in, the error introduced in the process is negligible.
For spectral analysis, spectra were extracted from each individual exposure
separately, then a joint fit was performed.

\section{SPATIAL ANALYSIS}
\label{s3}
\begin{figure}[!ht]
\epsscale{0.65}
\plotone{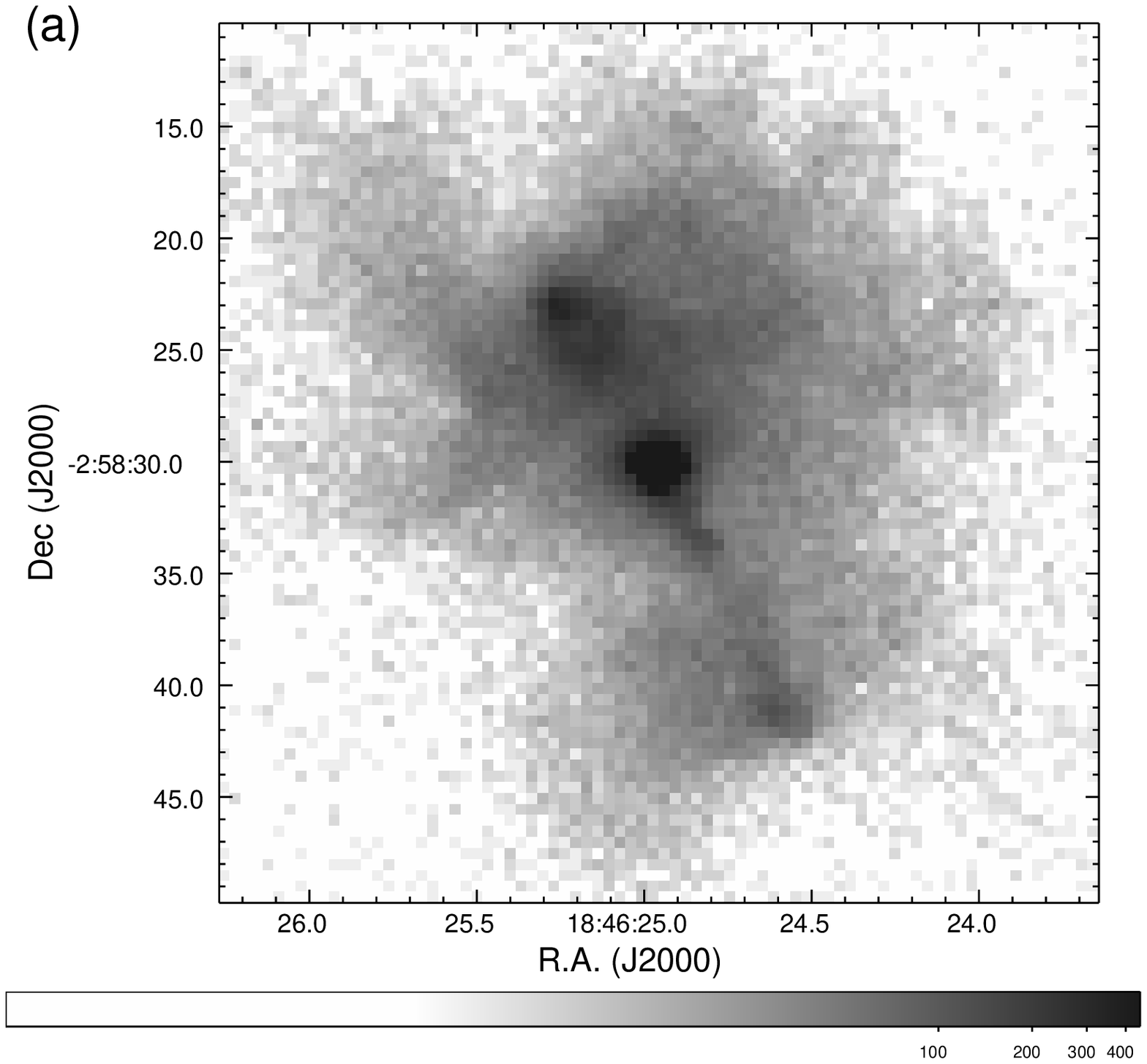}

\epsscale{1.1}
\plottwo{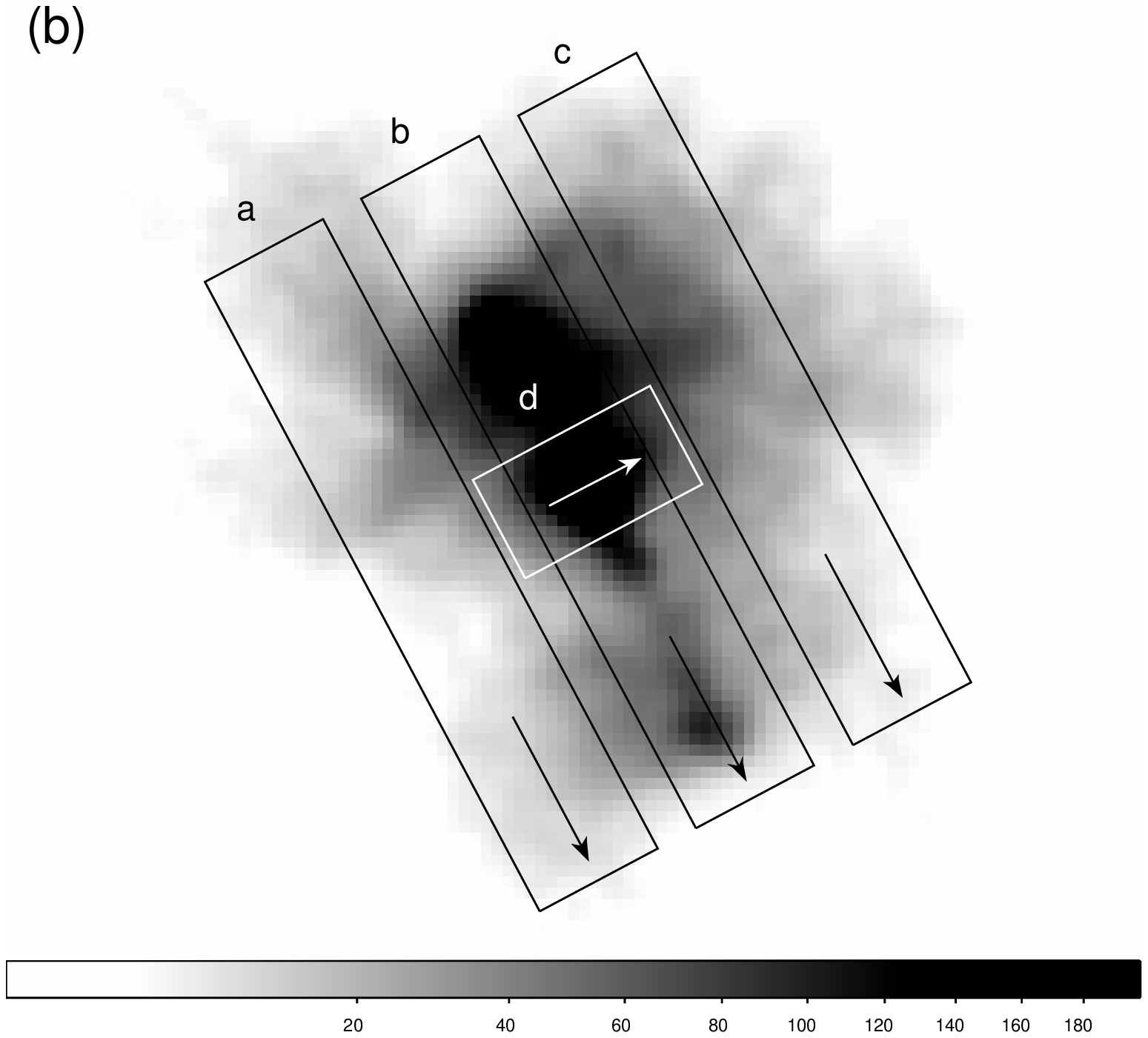}{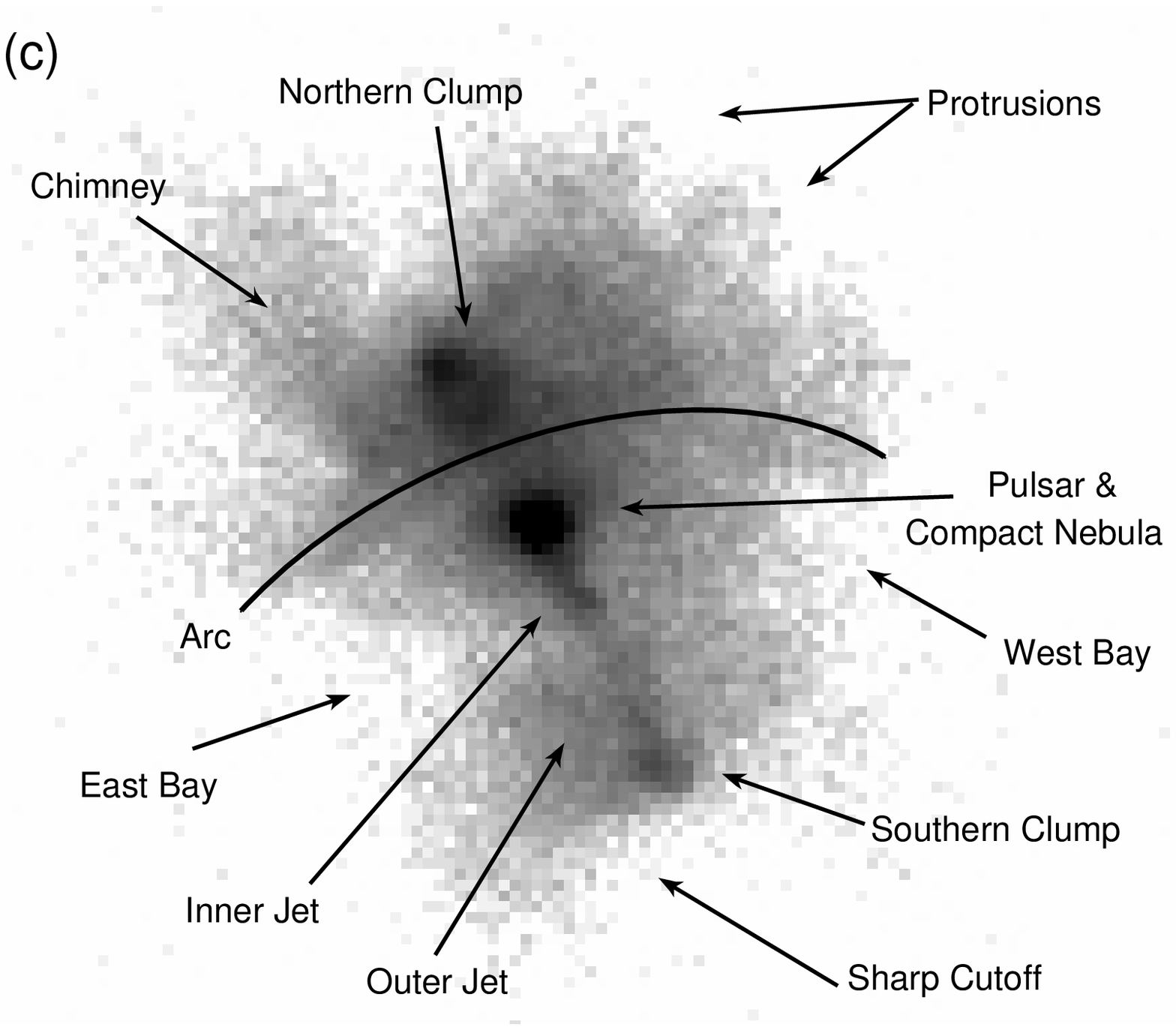}
\caption{(a) A Chandra ACIS-S image of the PWN in Kes 75 in the 1-8\,keV energy
range from the 2006 observation. (b) The same image lightly smoothed with
1\arcsec\ FWHM Gaussian, and shown in a different stretch to emphasize the faint structure.
The boxes show the extraction regions for the count profiles in Fig~\ref{f2}, with
the arrows indicating the positive $x$ direction in the plots.
(c) Small-scale features in the PWN.
\label{f1}}
\end{figure}

\begin{figure}[!ht]
\epsscale{1.0}
\plottwo{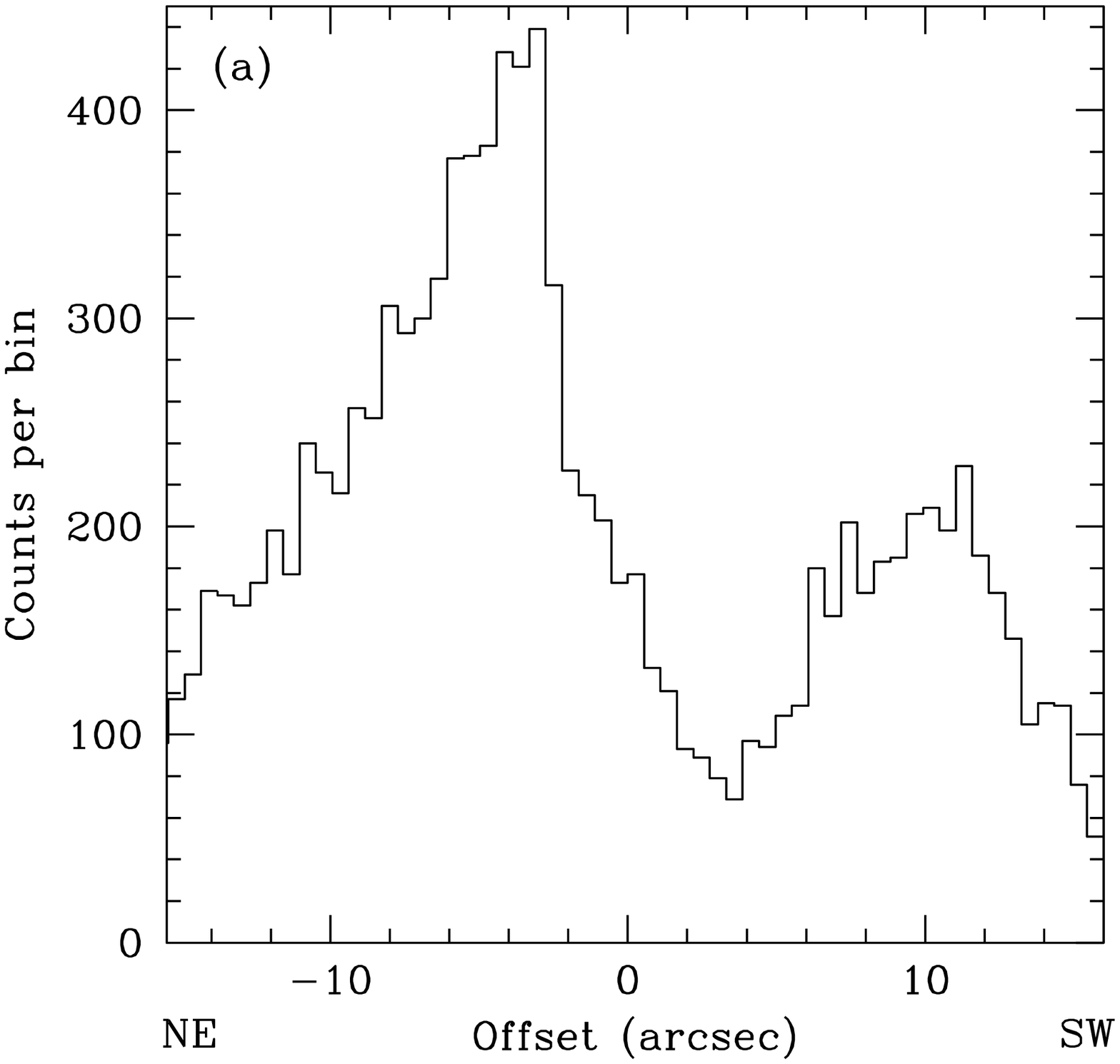}{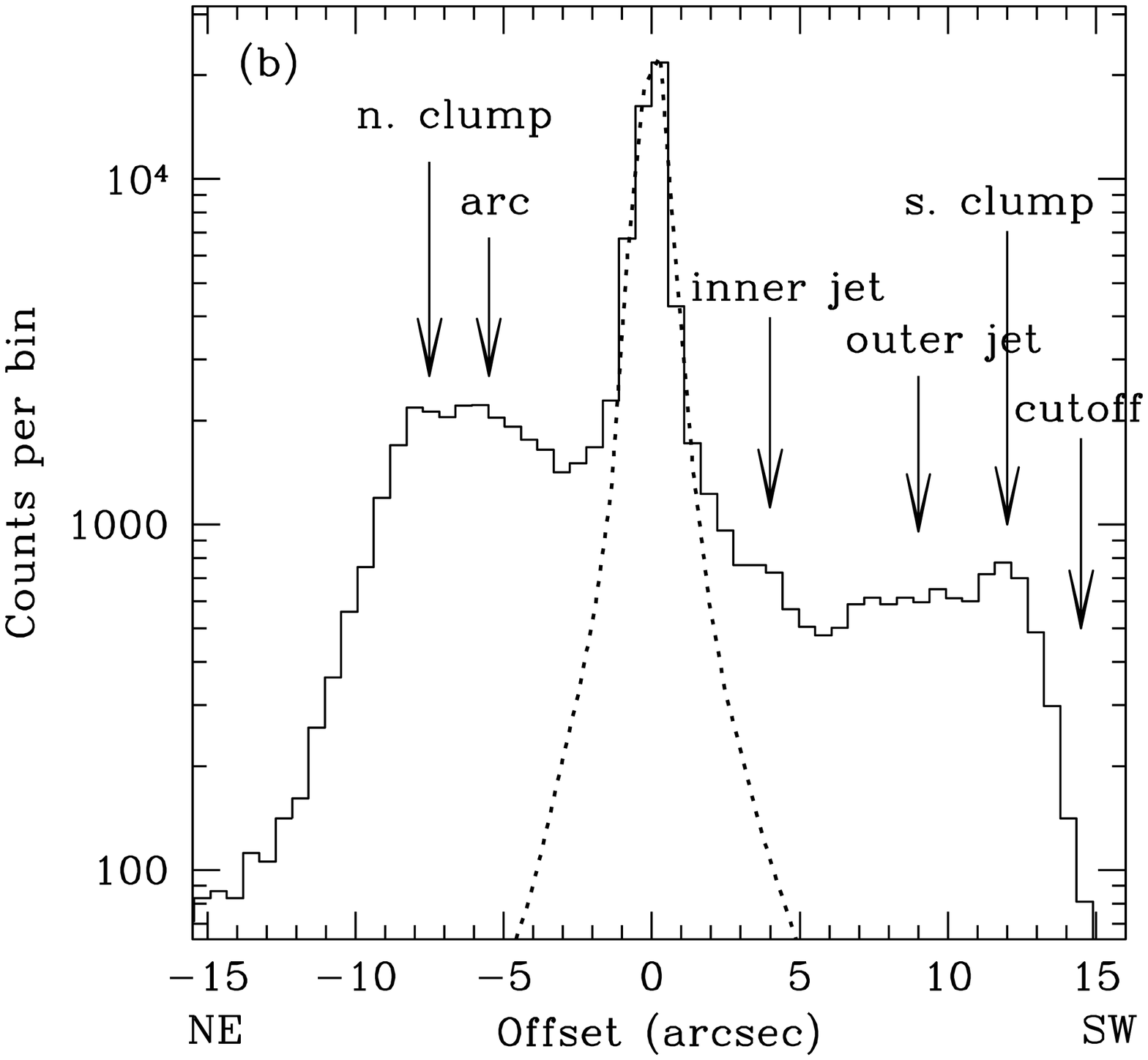}

\plottwo{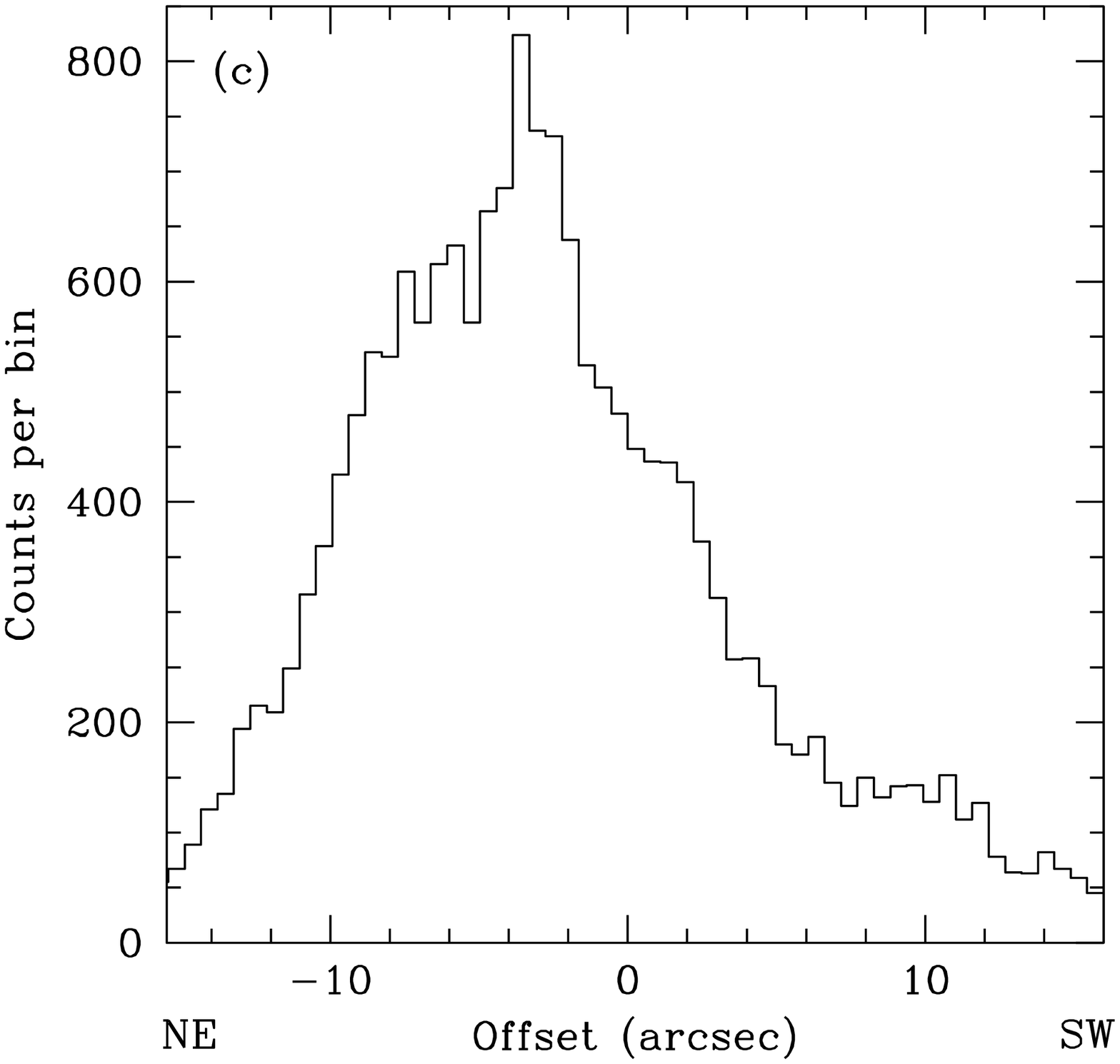}{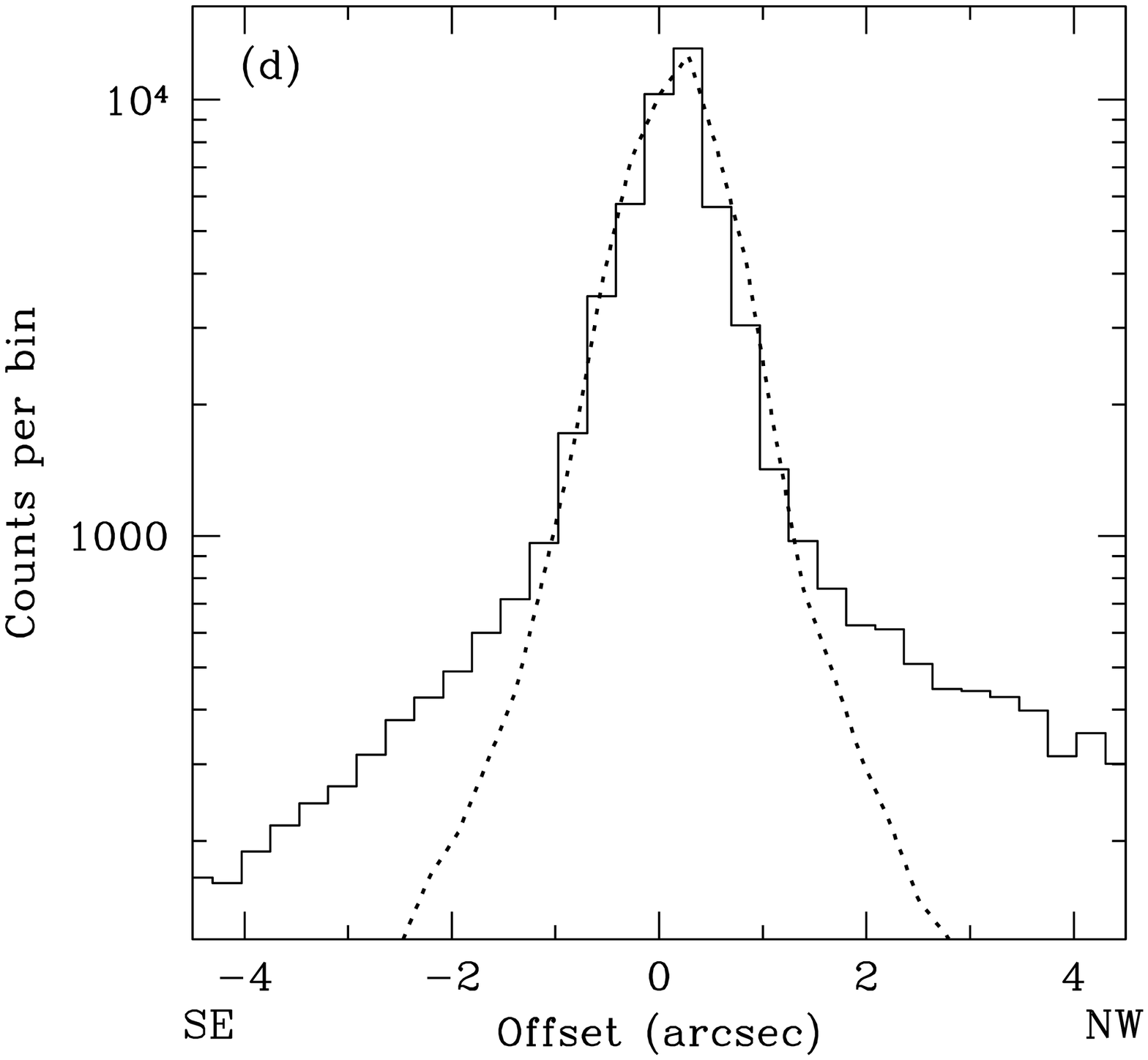}
\caption{Count profiles along different regions pulsar in Fig.~\ref{f1}, extracted from
the 2006 data in 1-8\,keV energy range. The $x$-axis is the offset from the
pulsar position, in arcseconds. The solid and dotted lines corresponding to the data and
the model PSF respectively. The latter is generated using the source spectrum with CCD
pile-up effects included. Panel (b) illustrates the features along the nebula symmetry
axis, while panel (d) shows the compact diffuse emission surrounding the pulsar. The
peaks in panels (a) \& (c) clearly illustrate the arc-like feature in the north.\label{f2}}
\end{figure}

Fig.~\ref{f1} shows the complex morphology of the PWN. As reported previously
by \citet{hcg03}, the $\sim30\arcsec$ diameter overall structure is filled with
diffuse emission and elongated in the northwest-southeast direction with an obvious
symmetry axis. The nebula has a sharp edge in the southeast, but a more diffuse
boundary in the north with finger-like protrusions. The PWN shows small-scale
features that resemble those of the Crab nebula \citep{stf06}, including bays in
the east and west, and a chimney in north. At the center of the PWN, a bright point
source at the pulsar position is embedded in a compact nebula and has a one-sided
collimated jet-like feature extending $5\arcsec$ to the southwest along the nebula
axis. This `inner jet' then fades and connects to a broader `outer jet' before it
terminates at a clump $13\arcsec$ southwest of the pulsar. The whole PWN is also
cutoff sharply at that point. In the northeast, there is another bright clump about
$6\arcsec\times4\arcsec$ aligned with the jet direction. At the base of this
northern clump, an arc-like feature, which runs across the entire nebula, is
bisected by the jet. Fig.~\ref{f1}c illustrates all these features and the count
profiles of the PWN along different regions are plotted in Fig.~\ref{f2}. In
particular, the peaks in Fig.~\ref{f2}a \& \ref{f2}c clearly indicate the arc-like
feature.

\subsection{SPATIAL MODELING}
\label{s31}
Highly collimated polar jets and equatorial outflows such as tori or wisps in
X-ray are commonly found features among young PWN systems \citep[see, ][]{gs06,kp08}.
Magnetohydrodynamics (MHD) simulations show that a latitude-dependent pulsar
wind can lead to Doppler-boosted arcs and jets in the X-ray synchrotron emission
\citep[e.g.][]{delz06}. Employing a simple model to capture these characteristic
features, \citet{nr08} demonstrate a robust way to measure the underlying PWN
geometry. For the PWN in Kes~75, the highly collimated jet-like feature and its
alignment with the nebula symmetry axis strongly suggest its connection to the
pulsar polar outflow. In this scenario, the arc-like feature perpendicular to
the jet could be part of a Doppler boosted torus or wisp in the equatorial plane.
We have applied the pulsar wind torus fitting scheme described by \citet{nr08} to
model the geometry of these features. Our model consists of a point source, a torus
and three Gaussian ellipses that represent the compact diffuse nebula surrounding
the pulsar, and the jets and clumps on both sides. An energy weighted point
spread function (PSF), which is simulated with the Chandra Ray Tracer (ChaRT)
and MARX software packages using the source spectrum, was used in the fitting.
The PSF also includes the CCD readout trail and the pile-up effects. The ellipses
are fixed in size and oriented along the torus symmetry axis, while their intensity
and separation are determined by the fit. Additional model parameters include the
torus axis position angle $\Psi$ (from north to east), inclination angle $\zeta$
between the torus axis and the observer line of sight, radius $R$, blur $\delta$
for the torus profile, post-shock flow velocity $\beta$, and the intensity of the
point source. The best-fit parameters are determined by maximizing a Poisson-based
likelihood function. More details on the modeling and fitting can be found in
\citet{nr08}. Note that we performed the fit to the 2006 epoch only, in order
to avoid any complication from different PSFs and instrumental response between
observations.

\begin{figure}[!ht]
\epsscale{0.9}
\begin{center}
\plottwo{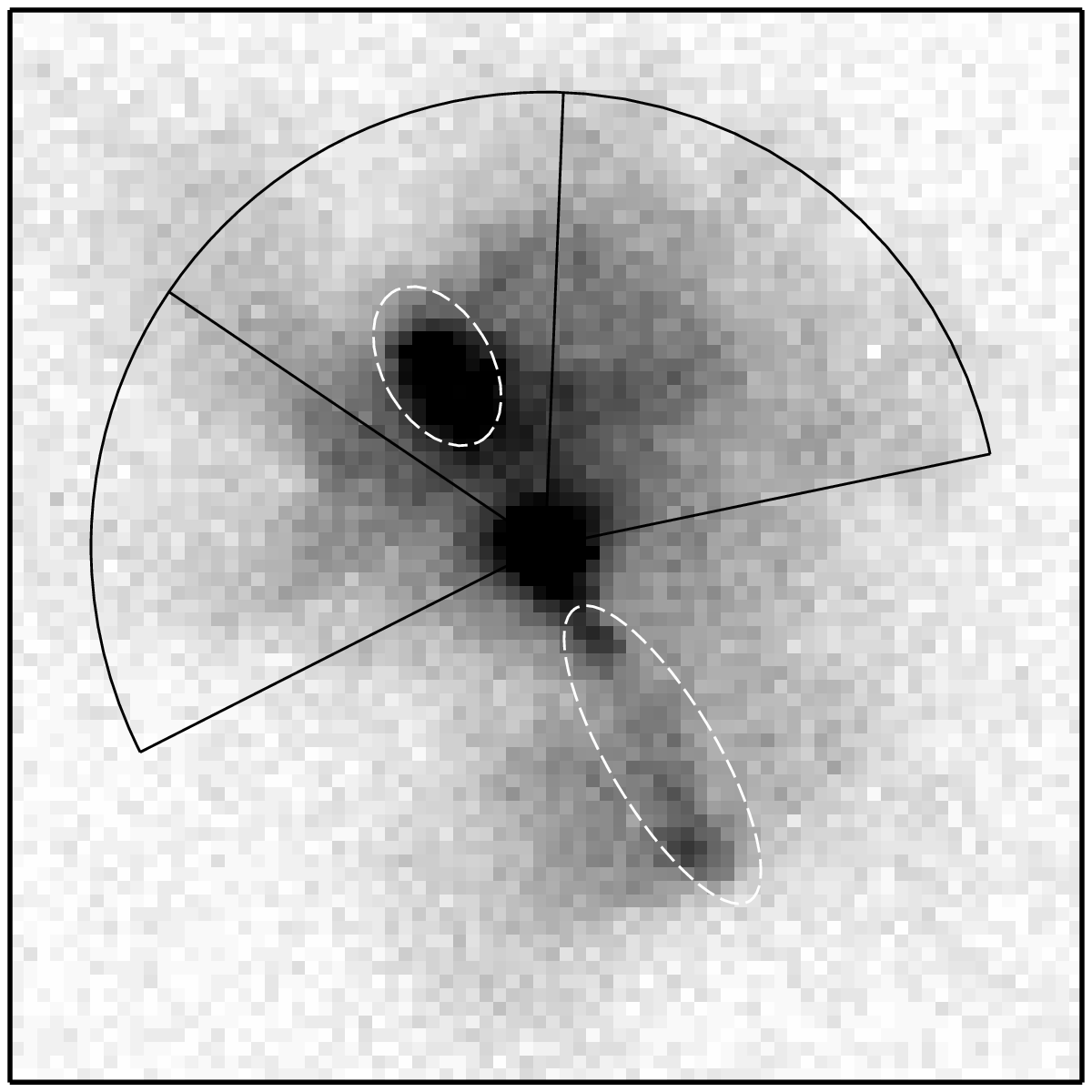}{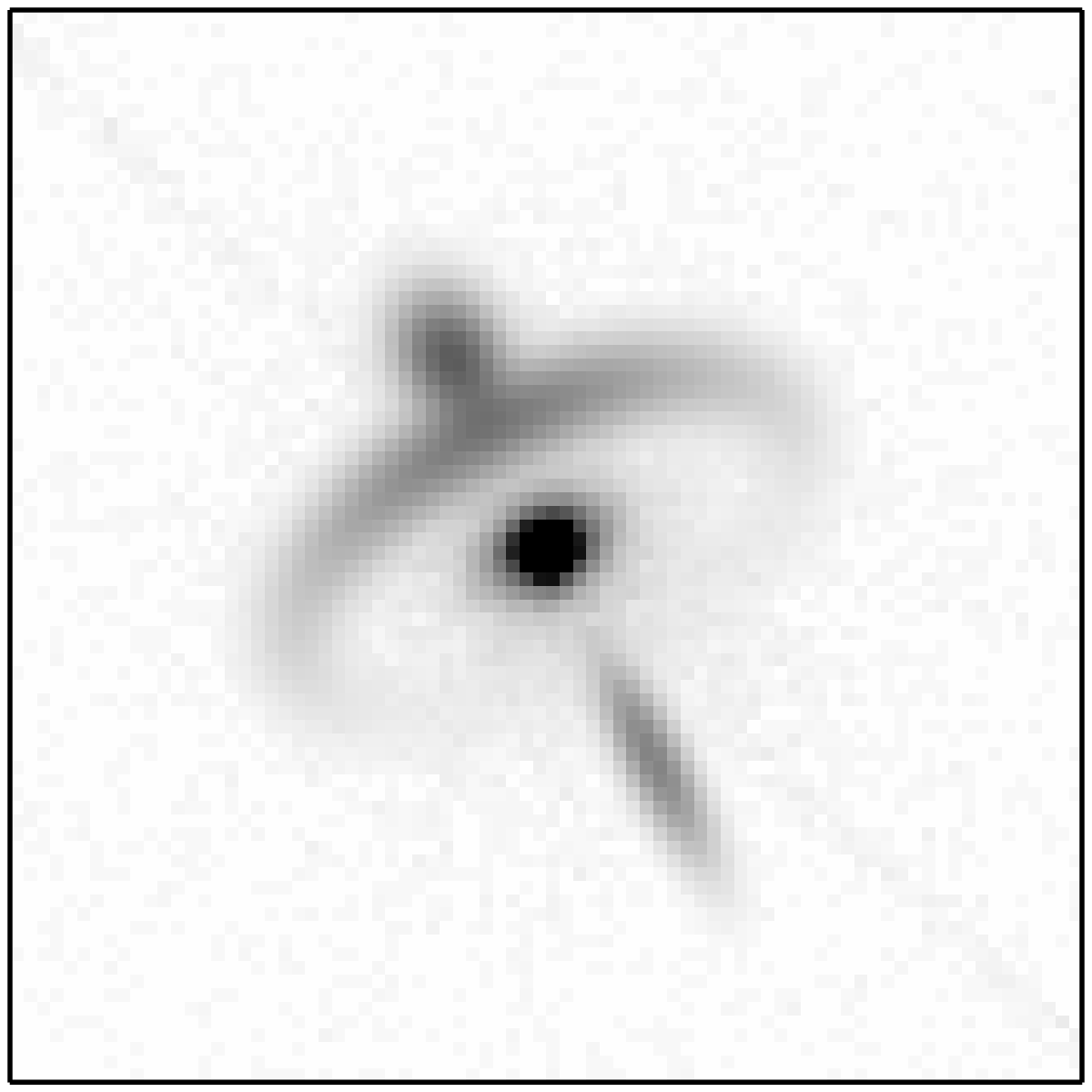}
\end{center}
\caption{\emph{Left:} same as Fig.~\ref{f1}a, showing the regions excised for
systematic errors estimate in the torus fitting (see text).
\emph{Right:} best-fit torus+jet model to the PWN structure.\label{f3}}
\end{figure}

Fig.~\ref{f3} shows the best-fit model compared to the data, with the
corresponding parameters listed in Table~\ref{tab1}. We found that the
uncertainties in the fit are dominated by the systematic errors, while
the statistical errors are negligible since the source is bright and the
deep exposure provides a high signal-to-noise ratio image. It is obvious
that our simple model captures only the characteristic features of the
PWN, hence the unmodeled structure contributes to the systematic
uncertainties. Following \citet{nr08}, we attempted to quantify this term
by blanking out the jet regions in the fit (the dotted ellipses in
Fig.~\ref{f3}). In addition, we also tried fitting only part of the arc
by excising the sectors in Fig.~\ref{f3}, one at a time. The change in the
parameters indicates how sensitively the results depend on a particular
region, therefore providing an estimate of the systematic errors in the fit
which are listed in Table~\ref{tab1}. At a distance of 6\,kpc, the best-fit radius
of the arc has a de-projected physical size of 0.3\,pc, which falls between
values for the inner and outer tori of the Crab pulsar \citep{nr08}, is a
reasonable value for a young PWN. The fitting results also indicate a
substantial flux contribution from the compact nebula, and that the northern
clump is brighter than the southern one. Given the inclination of the pulsar
spin axis determined by the fit, a simple picture of Doppler boosted jets
predicts the southern clump to be brighter than the northern one, which is
clearly inconsistent with the results.

\section{VARIABILITY}
\label{s4}
\begin{figure}[!ht]
\begin{center}
\includegraphics[width=0.9\textwidth]{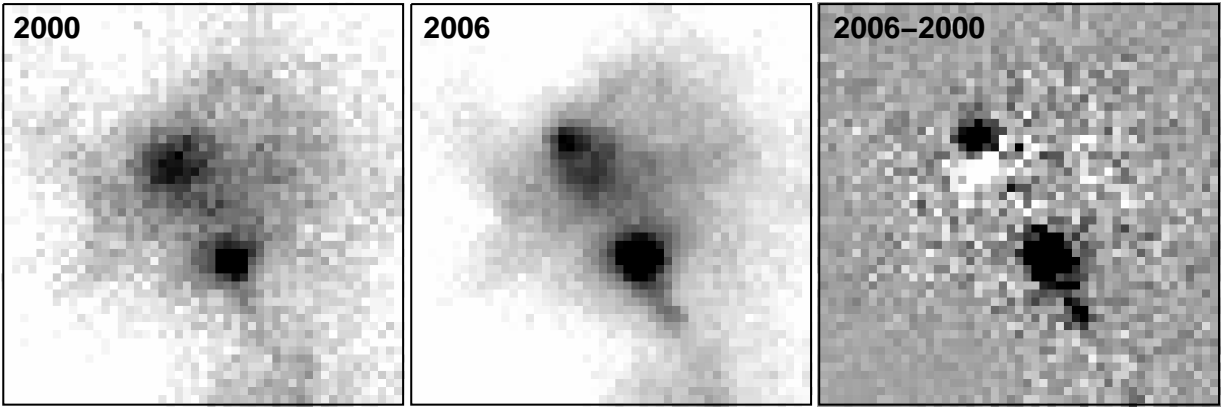}
\end{center}
\caption{Exposure corrected images of the PWN in Kes 75 in the 1-8\,keV band,
generated with weighted exposure maps using the best-fit PWN spectrum. The first
and second epoch are shown in the left and middle panels respectively with log
scales ranging from 0 to $10^{-4}$\,cts\,cm$^{-2}$\,s$^{-1}$\,pixel$^{-1}$.
The right panel shows the difference when the 2000 image is subtracted from the
2006 image. This greyscale is linear, ranging from $-10^{-6}$ to
$10^{-6}$\,cts\,cm$^{-2}$\,s$^{-1}$\,pixel$^{-1}$. \label{f4}}
\end{figure}

The 2006 exposure was started 7 days after the first burst reported by
\citet{gav08} and no bursts have occurred during the observation. The pulsar
lightcurve shows no short-term variability in and between the four individual
exposures. On the other hand, a comparison to the 2000 exposure reveals a
drastic increase in the pulsar flux. Fig.~\ref{f4} shows the exposure corrected
images of the PWN for the two observations. In the second epoch,
the pulsar is so bright that the CCD pile-up severely complicates the count rate
measurement. Therefore we estimated the flux from the CCD readout trail which is
due to the `out-of-time images' during the $40\mu$\,s frame transfer time
from one row to another. We employed a rectangular region of 3\arcsec\ width
centered on the trail, and selected regions far away from the PWN to avoid
any contamination. We tried different background regions and obtained a net count
of $2.7\pm0.3$\,cts\,row$^{-1}$ in 1-8\,keV range for the 2006 observation. The
uncertainty comes from both the statistical fluctuation and systematic errors due
to the background estimate. We tried independent measurements for the northeast
and southwest trails, as well as for each individual exposure, and the results are
all consistent. Since the readout trail consists of not only counts from the pulsar,
but also those from the PWN and SNR background in the same columns, the latter have
to be subtracted out. Excluding the central 3\arcsec\ radius from the pulsar, there
are $21\times10^3$ total counts (1-8\,keV range) in a 3\arcsec\ wide rectangular
region covering the whole trail of 600 CCD pixels in length. Of these counts, the
above estimate suggests that only $1.5\times10^3$ counts are from the readout
artifact and the rest are real events from the PWN, SNR and background. Therefore,
for a frame time of 1.8\,s, the PWN and SNR contribute a total of
$0.4$\,cts\,row$^{-1}$ to the readout trail, implying an unpiled count rate of
$0.7\pm0.1$\,cts\,s$^{-1}$ for the pulsar in 1-8\,keV range. Note that this is
only an upper limit of the true pulsar flux, due to the contamination by the
surrounding compact nebula. This is discussed in the next section.
Repeating the same exercise on the first epoch obtains a pulsar count rate of
$0.2\pm0.1$\,cts\,s$^{-1}$. If the pulsar spectrum and the luminosity of the
compact nebula remain unchanged between the epochs, the pulsar flux has to
increase by a factor of 3.5 in the 1-8\,keV energy band.

The total flux of the PWN shows no physical changes between the epochs. Excluding
the central 3\arcsec\ radius, the overall PWN has count rates of 0.59 and
0.61\,cts\,s$^{-1}$ in 1-8\,keV band for the first and second epochs respectively.
The 3\% difference is entirely due to the scattering of pulsar counts in the
same field-of-view. We generated a model PSF without any pile-up effects, and found
that the above region contains 3.5\% of the point source counts. Therefore, the
increase in the pulsar luminosity can easily explain the excess counts in the
PWN.\footnote{(0.7--0.2)\,cts\,s$^{-1}\times 3.5\%\approx0.02$\,cts\,s$^{-1}$}

\begin{figure}[!ht]
\epsscale{0.5}
\plotone{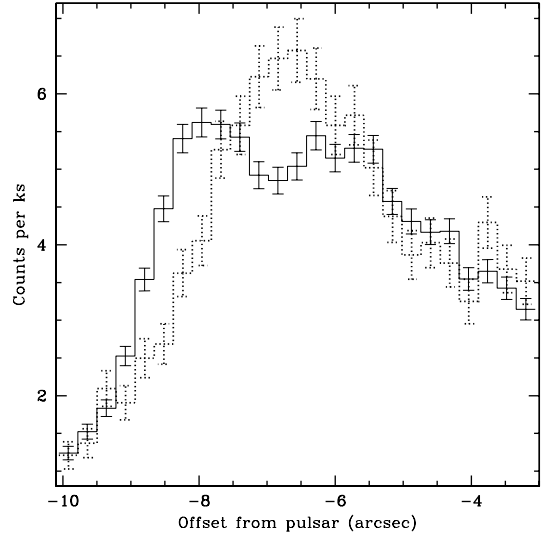}
\caption{Count profiles of the northern clump in the PWN of Kes 75, extracted in
a 4\arcsec\ wide rectangular region along the jet direction in the 1-8\,keV range.
The first and second epochs are shown by the dotted and solid lines respectively.
\label{f5}}
\end{figure}

On the other hand, some small-scale features in the PWN do demonstrate significant
variability between the observations. The exposure corrected images in Fig.~\ref{f4}
clearly show the evolution of the northern clump and the brightening of the southern
inner jet. In the 2006 data, the northern clump broadens and shows a double peak
feature. The count profiles in Fig.~\ref{f5} indicate that its peak has shifted
2\arcsec\ outward, while the total flux remains similar. The southern inner jet also
shows hints of morphological evolution: there is an obvious eastern extension at the
end of the jet in the 2006 exposure, but it seems to be absent in the first epoch.
Detailed studies of the flux and spectral variations of these features will be
presented in the following section.

\section{SPECTRAL ANALYSIS}
\label{s5}
\subsection{THE PWN SPECTRUM}
\begin{figure}[!ht]
\epsscale{0.9}
\plotone{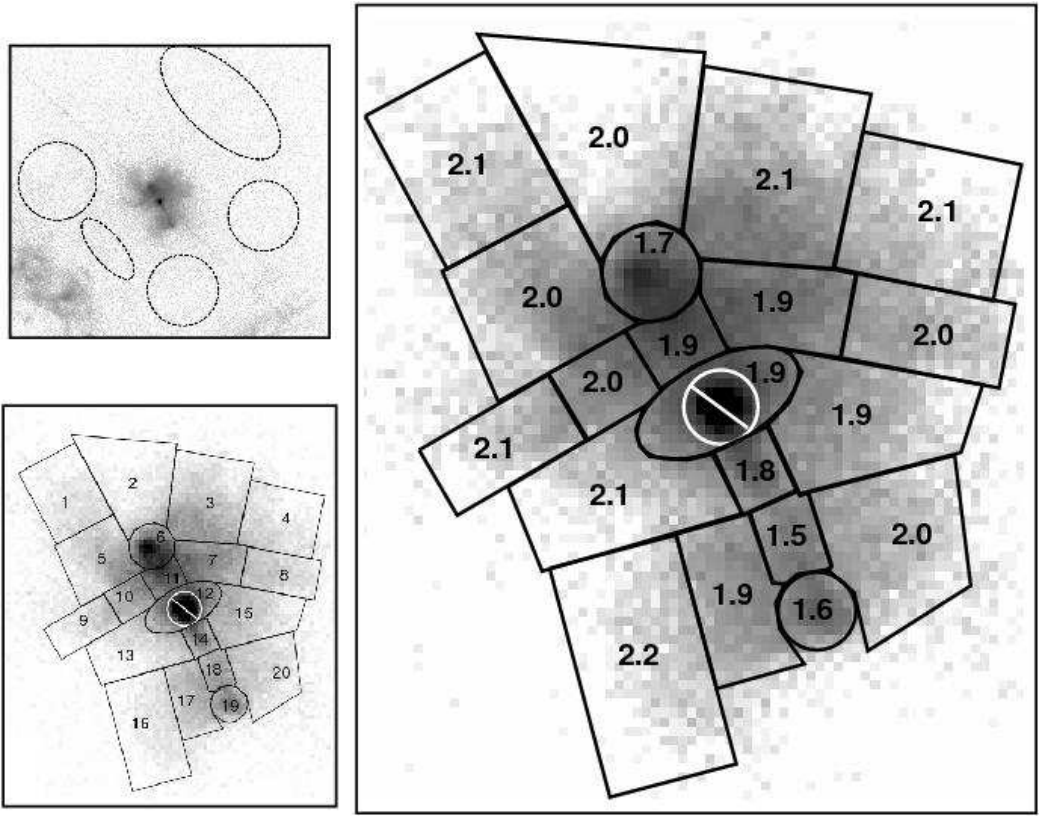}
\caption{\emph{Top left:} background regions used in the overall PWN and pulsar
spectral analysis. \emph{Bottom left:} extraction regions for the PWN spectrum.
\emph{Right:} best-fit powerlaw indices for the 2006 observation.\label{f6}}
\end{figure}

To investigate the spatial variation of the PWN spectrum, we divided the PWN into
different regions and extracted their spectra using the script {\tt specextract} in
CIAO. The source and background regions are shown in Fig.~\ref{f6}. As mentioned,
the spectra for the 2006 observation were extracted from the four exposure individually,
and then fitted jointly. All the spectral fits in this study were carried out in the
0.5-9\,keV energy range using the SHERPA environment. The spectra from different
regions were fitted simultaneously with a global column density but different powerlaw
indices. Careful examination of each spectrum shows no spectral or flux variation
between the exposures in the 2006 observation. The best-fit results are listed in
Table~\ref{tab2} with the corresponding uncertainties in the parameters. The
uncertainties reported here, as well as for the rest of the paper, are 90\% confidence
intervals for the projected multi-dimensional values, except for the flux values,
which are 1-D errors. The best-fit column density
$N_{\rm H}=(4.06\pm0.07)\times 10^{22}\,$cm$^{-2}$ obtained from the 2006
observation is consistent with the previous measurements \citep{hcg03,mor07,ks08}.
Comparing the total flux between the epochs indicates a 3\% difference, which is in an
excellent agreement with the count rate estimate in the above section.

To determine any spectral variations in the overall PWN between the epochs, we
refitted both data sets with the column density fixed at $N_{\rm H}=4.0
\times10^{22}\,$cm$^{-2}$, in order to avoid any possible covariance between
$N_{\rm H}$ and other spectral parameters. The results are shown in Table~\ref{tab2},
indicating no systematic changes in the spectral indices between the observations,
and the fluctuation is mostly within the measurement uncertainties. On the other
hand, the table suggests flux variations in regions 1,2,6,9,12, and 14. In particular, the change
in region 6 is introduced by the shift of the northern clump into the region. For region
12, the apparent flux increase is the result of pulsar counts scattered into the region,
rather than physical changes. From the same analysis as the above section, we have worked
out the expected pulsar flux in the region and found that the value is consistent with the
excess flux observed. Moreover, the same exercise on the other regions indicates that
the flux variations are likely to be physical, but we note that the total flux of the PWN
remains unchanged.

\begin{figure}[!ht]
\epsscale{0.5}
\plotone{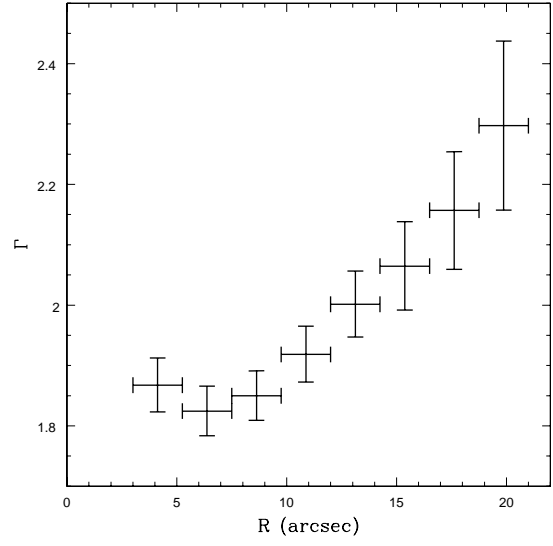}
\caption{Variation of the PWN photon index in Kes 75 with the radius from
the central pulsar.\label{f7}}
\end{figure}

Fig.~\ref{f6} shows the spectral map of the PWN. The compact nebula at the center
has a relatively soft spectrum as compared to the innermost part of many other young
PWNe. Also, regions along the symmetry axis are systematically harder than the surrounding
emission, and the overall spectrum softens as one progresses to larger scales.
To further investigate this variation, we extracted the PWN spectra in annular
regions and fitted jointly to different powerlaws with a global $N_\mathrm{H}$.
The result for the 2006 data is plotted in Fig.~\ref{f7}, the one from the 2000 data
is very similar. Except for the innermost bin, the spectral index increases monotonically
with radius. This trend is commonly observed among PWNe, including the Crab, 3C~58
and G21.5--0.9 \citep{mor04,sla04,sla00,war01}.

\begin{figure}[!ht]
\begin{center}
\includegraphics[width=0.5\textwidth]{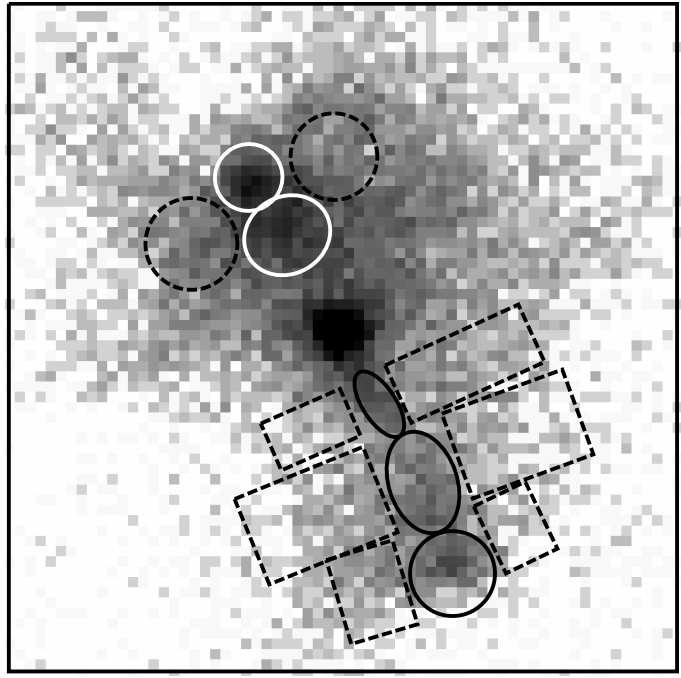}
\end{center}
\caption{Extraction regions for the PWN features shown by the solid regions
from north to south: northern clump 1, northern clump 2, inner jet, outer jet,
southern clump. The corresponding backgrounds are extracted from the adjacent
regions shown by the dotted lines. The underlying image is the 2006 data
in the 5-9\,keV energy band, showing high energy emission in the PWN.\label{f8}}
\end{figure}

In addition to the overall nebula, we also carried out spectral analysis for the
individual PWN features. To allow a direct comparison between the structures, 
we minimize the nebular contamination by subtracting the background from adjacent
regions. The extraction regions for the northern clump, southern clump, inner jet,
outer jet and southern clump are shown in Fig.~\ref{f8}. As the northern clump shows
a double peak structure in the 2006 observation, we further divided the extraction
region to determine any spectral difference from the two peaks. The best-fit spectral
parameters with 90\% confidence intervals are listed in Table~\ref{tab4}. The spectral
properties vary substantially between the structures. To check if this could be caused
by a variable column density, we tried fitting $N_\mathrm{H}$ in the 2006 observation.
However, the lower signal-to-noise ratio of the 2000 data precludes a similar exercise.
The outer jet seems to have a harder spectrum than the other regions, if it is real, it
could indicate either high energy emission or a larger absorption column density in
the region. For the northern clump, we do not find any spectral difference for its two
peaks. Also, the powerlaw index and total luminosity of the clump show no significant
change between the two epochs. On the other hand, the flux of the inner jet doubles
in the 2006 observation, which could not be accounted for by the scattering of the
pulsar PSF, since the extraction region is small. Variability in pulsar jets has been
observed among many PWN systems, such as the Crab, Vela and B1509--58, with timescales
from days to years \citep{hes02,pav03,del06}. Therefore it is not unexpected to
detect variation in the inner jet here, given the two epochs are separated by 6 years,
see \S\ref{s62} for more discussion. We have tried fitting the clump spectra with
thermal models to investigate the scenario in which the clumps represent a shock
interaction between the jets and the surrounding materials. We found that a thermal
bremsstrahlung provides the best fit among the thermal models, since there is no obvious
emission line feature in the spectra. However, as listed in Table~\ref{tab4}, the
temperature is not very well constrained and statistically the fit is slightly worse
than a simple powerlaw. We are thus inclined to infer the emission as non-thermal
as seen in other PWNe.

\subsection{THE PULSAR SPECTRUM}
Although the readout trail of the pulsar is free from pile-up, it is much fainter
than the background emission from the PWN and the SNR shell to yield any high
quality spectrum for the pulsar. Therefore, the pulsar spectrum was extracted 
from the central 2\arcsec\ radius using the script {\tt psextract} with the same
background regions shown in Fig.~\ref{f6}. We rebuilt the response matrix files
(RMFs) with the tool \texttt{mkacisrmf} which corrects for the CTI, and employed
the \citet{dav01} model in the fitting to account for the CCD pile-up effects.
Again, for the 2006 observation, the spectra were extracted from individual exposure 
and then fitted jointly. We did not attempt to fit the column density, as this
could be complicated by the pile-up effects and by any possible thermal emission
from the pulsar. Instead, it was fixed at the PWN value $N_{\rm H}=4.0\times10^{22}
\,$cm$^{-2}$ in the analysis. Also, due to contamination from the surrounding compact
nebula, a simple background subtraction with an annuli region may lead to inaccurate
pile-up estimate. Hence, we fit the source spectrum with a two-powerlaw model: one
for the background and the other one for the pulsar. The former is fixed at the
best-fit spectrum of region 12 (see Fig.~\ref{f6}) and we scale its amplitude
according to the spatial modeling in \S\ref{s31}. We found that the compact nebula
provides 0.05\,cts\,s$^{-1}$ (1-8\,keV range) to the extraction aperture and this
is fixed as the nebular background in the fits. This spectral model provides an
adequate fit to the 2000 data and yields a spectral index $\Gamma=1.1\pm0.1$ and
an unabsorbed flux of $6.1\times10^{-12}$\,ergs\,cm$^{-2}$\,s$^{-1}$ in the
0.5-10\,keV range. The nebular background contributes $\sim20\%$ of the observed
flux and the overall spectrum has a pile-up fraction less than 10\%.

For the 2006 observation, we started with the same two-powerlaw model, but found
that the pulsar spectrum is very close to that of the surrounding compact nebula.
Therefore in the final fit, only a single absorbed powerlaw model was used.
Table~\ref{tab3} lists the best-fit spectral parameters, indicating a much softer
spectrum ($\Gamma=1.86\pm0.02$), with a factor of 6 increase in the unabsorbed
flux ($3.4\times10^{-11}$\,ergs\,cm$^{-2}$\,s$^{-1}$) in the second epoch. Due to
spectral softening, the observed flux only changes by a factor of 4, suggesting
a 6\% nebular contamination and a 20\% pile-up fraction despite a shorter frame
time. Comparing the individual spectrum in the 2006 observation shows no
short-term flux or spectral variation for the pulsar at this time. As a check for
consistency, we convert the above results to the un-piled count rates for the
pulsar plus the nebular background. The inferred count rates in the 1-8\,keV band
are 0.17 and 0.77\,cts\,s$^{-1}$ for the 2000 and 2006 observations respectively.
This is in good agreement with the independent estimate from the readout trails in
the previous section. We also note that our results are mostly consistent with the
values reported by \citet{ks08} and \citet{gav08}, but the pulsar spectral index
for the first epoch is slightly lower than the value reported by \citet{hcg03}.
The difference is probably the result of improvements in the pile-up model.

In order to look for any thermal emission from the neutron star, we have tried
adding a blackbody component to the fits. As shown in Table~\ref{tab3}, the PL+BB
model provides a better fit to the 2006 data in terms of the reduced $\chi^2$ value.
Moreover, an $F$-test indicates that this model is statistically better than a single
powerlaw at a 99.97\% confidence level. Hence, we concluded that the thermal
component is significant. The best-fit blackbody has a temperature of $0.9\pm0.2$\,keV
and a flux $\sim10\%$ of the non-thermal emission. This implies a blackbody radius
of $R^{\infty}=0.4\pm0.2$\,km at 6\,kpc. For the first epoch, a thermal component
is not observed and we place a detection limit $4\times
10^{-13}$\,ergs\,s$^{-1}$\,cm$^{-2}$ at 90\% confidence level in 0.5-10\,keV energy range.

\section{DISCUSSION}
\label{s6}
\subsection{THE ENVIRONMENT AND STRUCTURE OF THE PWN}

At a distance of 6\,kpc, the overall PWN has a radius
$R_\mathrm{PWN}\approx 20\arcsec=0.6$\,pc and an X-ray luminosity
$L_X^{0.5-10}=1.4\times 10^{35}$\,ergs\,s$^{-1}$ in the 0.5-10\,keV
band. A very rough estimate on the nebula magnetic field can be made based on
the X-ray luminosity with the equipartition assumption, or from the SNR
evolution assuming a constant pulsar spin-down power \citep{sew84}. They both
suggest a field strength $B\sim40\,\mu$G.  On the other hand, comparing the PWN
X-ray and radio spectra \citep{hcg03,bg05} indicate a spectral break at
$\nu_b\approx 5\times 10^{14}\,$Hz \citep{mor07}. If this is solely due to
synchrotron cooling, then the pulsar spin-down of $\sim800$\,yr implies a nebula
field of $B\sim100$\,mG. As \citet{mor07} point out, this is likely to be too
large for the PWN. \citet{bg05} also suggest that the simple standard synchrotron
cooling scenario may not hold for this object. Indeed there are many intrinsic
and evolutionary effects that could complicate the picture \citep[e.g.,][]{rc84, sla08}.

The HESS detection of VHE $\gamma$-rays from Kes~75 provides a better handle on
the magnetic field \citep{dja08}. In a one zone model assuming all the TeV photons
come from inverse Compton scattering with the cosmic microwave background (CMB),
the X-ray and $\gamma$-ray fluxes are related to the energy densities of the
magnetic field ($\epsilon_B$) and the photons ($\epsilon_{\rm ph}$) by
$L_X/L_\gamma=\epsilon_B/\epsilon_{\rm ph}$ where $\epsilon_{\rm ph}=3.8\times
10^{-13}$\,ergs\,cm$^{-3}$ for the CMB. Comparing our result of X-ray flux
in 0.5-10\,keV band to the reported HESS detection of TeV flux in 0.3-5\,TeV
range suggests $L_X/L_\gamma\sim 30$, which is comparable to that of G21.5--0.9,
implying a low magnetic field of $B\approx15\mu$\,G \citep{dja08}. The picture is
further supported by the similar radio and X-ray PWN sizes. As compared to the
radio images from the VLA and the BIMA array, the X-ray synchrotron emission
extends nearly to the edge of the radio counterpart \citep{hcg03,bg05}. This
suggests a long synchrotron cooling time scale and hence a weak magnetic field.
A nebula field $B=15B_{15}\mu$\,G is adopted throughout the discussion.\footnote{We
note that values between 10-40$\mu$\,G do not change our conclusions qualitatively.}
This field strength gives a magnetic pressure $10^{-11}B_{15}^2$\,dyne\,cm$^{-2}$
in the PWN, with a magnetic energy $2\times 10^{44}B_{15}^2$\,ergs assuming
the nebular geometry is a sphere. This is only 0.1\% of the total spin-down energy
if the pulsar was born with a spin period close to the current value; if the
pulsar's initial spin period is shorter, the fraction will be even smaller.

Numerous theoretical and numerical studies of PWN evolution within an expanding
SNR suggest that during the first $\sim1000$ years, a PWN expands freely into
the surrounding ejecta before the supernova reverse shock is reached
\citep[e.g.][]{van01,buc03,van03,che05}. Given the young age of Kes~75, the PWN
is likely to be in the free expansion phase. The expansion of light pulsar wind
materials into the dense supernova ejecta could result in Rayleigh-Taylor
instabilities. This may be associated to the protrusions in the PWN boundary,
similar to the optical filaments of the Crab Nebula \citep{hes96}. Alternatively,
the protrusions in the north could also due to ordered magnetic fields, as
\citet{stf06} proposed for the Crab PWN. Although a comparison to optical
observations will help to identify the nature of these structures, this may not
be observationally feasible due to the large distance of the source and its
location on the Galactic plane.

To interpret the arc-like feature in the northeast, we first note that it is
unlikely to be reconciled as a bow shock. Due to the young age of the system,
the pulsar is not expected to move supersonically within the SNR interior. 
Therefore we suggest that the arc could be a toroidal termination shock, which
is a commonly observed feature among young PWNe. The shock radius $r_s$ is
determined by the balance between the pulsar ram pressure and the ambient
pressure $P$: \[ r_s^2=\frac{\dot{E}}{4\pi c\xi P} \ , \] where $\dot{E}$ is the
pulsar spin-down luminosity and $0<\xi<1$ is the fraction of a sphere covered by
the pulsar wind (e.g.\ $\xi=2/3$ if the wind has a latitude dependence of
$\sin^2\theta$). In this picture, the inferred pressure for our PWN is $P\sim 10^{-11}
\xi^{-1}$dyne\,cm$^{-2}$, consistent with the magnetic pressure inside the
nebula. This is also in a general agreement with the numerical simulations of
a generic young PWN evolving in a SNR \citep{van01,buc03}. Comparing to the thermal
pressure in the SNR limbs \citep{mor07}, this value is $\sim100$ times lower,
confirming that the supernova reverse shock has not yet reached the PWN and
that most of the ejecta in the SNR interior are still cold.

In general, we do not expect any significant synchrotron emission inside the
termination shock. In this case, the compact nebula surrounding the pulsar
could be a result of particle injection from previous outbursts. The
synchrotron cooling timescale for a particle emitting at $\varepsilon=5$\,keV in
the nebula is \[ t_{\rm synch}=39B^{-3/2}\varepsilon^{-1/2}\, \mathrm{kyr}
\simeq 300B_{15}^{-3/2}\,\mathrm{yr} \ . \] If an outburst occurs every decade,
as suggested by \citet{gav08}, then based on the timescale, this could possibly
power the emission of the compact nebula.

As an alternative, one could identify the compact nebula as a termination
shock. The smaller radius $r_s=3.5\arcsec$ only requires a slightly larger
magnetic field in the nebula and the ambient pressure is still much lower
than the thermal pressure in the rims. In this scenario, the arc-like feature
could be interpreted as an analogue to the outer arc in G320.4--1.2 powered by
PSR B1509--58, which \citet{gae02} argue is a wisp in the equatorial outflow
based on the particles flowing timescale. In our case, the flowing timescale
from the compact nebula to the arc-like feature is \[ t_{\rm flow} =
\frac{3r_s}{c}\left(\frac{r} {r_s}\right)^3 =30\,\left ( \frac{r_s}{3.5\arcsec}
\right)^{-2}\left( \frac{r}{10.1\arcsec}\right)^3\,\mathrm{yr} \ , \] which
is shorter than $t_{\rm synch}$. This indicates a different physical origin of
the arc than the Crab Nebula's bright outer torus, where most X-ray emitting
particles lose their energy. In contrast to the Crab Nebula's wisp \citep{hes02},
our data do not indicate any significant proper motion in the arc. Similar
results have also been reported for the arc surrounding PSR B1509--58
\citep{del06}, and the radio wisp in 3C 58\citep{bie06}. If the variability in
our case is compatible, it would be below the current detection sensitivity
given the relatively shallow exposure of the first epoch.

Finally, we discuss the spectral softening of the PWN with increasing distance
from the pulsar, as shown in Fig.~\ref{f7}. This is a result of particle energy
dissipation due to synchrotron burn-off and nebula expansion. Although some
theoretical models predict a rapid increase in the spectral index
\citep[e.g., fig. 4 in][]{sla04}, such degree of variation is not observed here. Indeed the
variation $\Delta\Gamma=0.5$ in our case is smaller than many other systems,
such as 3C~58 or G21.5--0.9 \citep{sla00,sla04}, indicating a long synchrotron
cooling time in the PWN. The slightly softer spectrum in the innermost bin could be
due to the injection of lower energy particles from any pulsar bursts in the past,
or simply because the next two bins are contaminated by the polar outflow which has a
harder spectrum.

\subsection{THE POLAR OUTFLOW}
\label{s62}
The jet-like features along the PWN symmetric axis show systematically
harder spectra than their surroundings, indicating a large velocity
in the polar outflow, i.e.\ $t_{\rm synch} > t_{\rm flow}$. If the absence
of a counter inner jet in the north is due to Doppler boosting, then its detection
limit provides an estimate of the flow speed. The flux ratio between the
inner jet and its expected counterpart is given by
\[ \frac{f_{\rm j}}{f_{\rm cj}} = \left ( \frac{c+v_j\cos\zeta}
{c-v_j\cos\zeta}\right )^{1+\Gamma} \] \citep{pel87}.
From the the best-fit inclination angle $\zeta=62\arcdeg$ and the photon index
$\Gamma=1.7$ for the jet, the detection limit on the counter jet sets its minimum
bulk velocity $v_j>0.4c$. This is comparable to jets in other pulsars
such as the Crab, Vela and B1509--58 \citep{hes02,pav03,gae02}. To estimate
the magnetic field in the jet, we approximate the southern inner jet by a
cylinder with 1.5\arcsec\ diameter and a length of $l_j=3.5\arcsec$, which
correspond to 0.04 and 0.1\,pc respectively at 6\,kpc. The best-fit spectral
luminosity $L_{\rm 5\,keV}=1.5\times 10^{31}$\,ergs\,s$^{-1}$\,keV$^{-1}$ at
5\,keV suggests an equipartition field $B_j\sim 120(1+\mu)^{2/7}\phi^{-2/7}\,\mu$G
with a total energy $E_j\sim5(1+\mu)^{4/7}\phi^{3/7}\times10^{43}$\,ergs for the
inner jet, where $\mu$ is the ion to electron energy and $0<\phi<1$ is the filling
factor \citep[see][]{gae02}. The synchrotron cooling time for particles emitting
at 5\,keV in the jet is then $t_{\rm synch}=13$\,yr, confirming that $t_{\rm synch}
\gg t_{\rm flow}\sim l_j/v_j=0.8\,$yr. This also implies a minimum power input
from the pulsar $\dot{E}_j\sim E_j/t_{\rm synch} = 1.2\times10^{35}$\,ergs\,s$^{-1}$,
which is 1.4\% of the total spin-down luminosity. 

We argue that the variability in the polar outflow is unlikely due to the recent
pulsar bursts. Since the 2006 observation was carried out 7 days after the bursts,
materials traveling from the pulsar to the end of the inner jet requires an apparent
superluminal motion $v_{\rm app}\approx 20\,c$ in the plane of the sky, which is related
to the true space velocity $\beta$ by $v_{\rm app}=\case{\beta\sin\theta}
{1-\beta\cos\theta}$, where $\theta$ is the angle between the direction of motion
and the observer line of sight. As $\beta\leq 1$, this places an upper limit $\theta\leq
5.7\arcdeg$. The best-fit pulsar spin axis inclination $\zeta=62\arcdeg \gg 5.7\arcdeg$
indicates that the observed variability in the jet could not be explained by any
superluminal motion in the polar outflow. For any emission in a random direction that
beamed towards the Earth with $\theta\leq 5.7\arcdeg$, there is an additional constraint
for the alignment with the jet direction on the plane of the sky. Therefore the chance
probability is very small and we conclude that the variability of the inner jet is almost
certainly unrelated to the recent bursts.\footnote{However, this does not completely
rule out the possibility that the variability is due to any previous bursts which
occurred years ago.} A similar argument also applies to the northern clump. Due to the
strong magnetic field and compact size of the inner jet, the Alfv\'en crossing time is
in the order of 100 days, which is comparable to $t_{\rm flow}$. Given the two epochs
separate by 6 years, we could not distinguish whether the inner jet variability is due to MHD
instability as is possibly so in the cases of Vela and PSR B1509--58 jets \citep{pav03,del06},
or simply due to materials flowing downstream.

As one moves downstream along the jet, spectral softening is generally expected as
a result of synchrotron burn-off. However, the outer jet region shows some hints of
a harder spectrum than the inner jet, which is difficult to understand unless there
is a continuous particle acceleration along the way \citep{hj88} or particle
injection from a shock (see below). An alternative explanation could be variability
in the absorption column density, although this requires a nearly 20\% change in
$N_\mathrm{H}$ across the regions (Table~\ref{tab4}). Further out along
the jet direction, emission on both sides of the pulsar terminates abruptly at the
northern and southern clumps. The similar spectral indices of the two clumps hint
that the emission may come from the same physical process, but their relative
brightness suggests that they could not be simply reconciled as Doppler boosted
jets. It is possible that the clumps are results of non-thermal emission from the
shock interaction between the polar outflow and the ambient medium. When the supersonic
pulsar jets encounter the supernova ejecta, the density gradient results in a bow shock
and leads to strong X-ray emission. Indeed the sharp edge of the PWN near the southern
clump provides some evidence of such a density gradient and this likely results in the clump.
The one in the north is less obvious, which could be due to projection effects. Comparing
to other PWN systems, a similar scenario has been proposed to associate the X-ray
features in RCW~89 with PSR B1509--58 \citep{tam96,yat05}. The difference in our case
is that the clumps are found on both sides and much closer to the pulsar. Besides, the
clumps are likely to be non-thermal. Also, due to the younger age of Kes~75, the jets
would encounter a cold ambient medium which has not yet been heated up by the supernova reverse
shock. If we compare to other astrophysical jets \citep[e.g.][]{hk06,deg05}, the clumps
may be identified as analogues to the terminal hotspots in the extragalactic jets,
which usually have non-thermal spectra well-modeled by synchrotron or synchrotron
self-Compton emission \citep{har01}. Then the outer jet in our case could be a backflow
cocoon formed by the postshock materials, and the hard spectrum in the region could be
related to the shock accelerated particles. Moreover, the variability of the northern
clump can be attributed to Rayleigh-Taylor instabilities, as the jet expands into the
much denser ejecta. The luminosity of the northern clump is $1.2\times
10^{34}\,\mathrm{ergs\,s^{-1}}=0.1\dot{E}_j$, which can be easily powered by
the jets, and the flow time scale $R_{\rm PWN}/v_j<20\arcsec/0.4c=5$\,yr is
substantially shorter than the synchrotron cooling time $t_{\rm synch}=13\,$yr.
Therefore the particles are still energetic on reaching the edge of the PWN.
Future multi-wavelength observations, in particular high resolution radio maps of
the PWN could help to further investigate the nature of the clumps.

\subsection{THE NEUTRON STAR}
The best-fit pulsar spectrum for the second epoch indicates a $\sim10\%$ flux
contribution from the blackbody component, which is not observed in the 2000 data.
This suggests that the thermal emission is likely related to the afterglow of the
recent bursts, and could be part of the transition between a rotation-powered pulsar
to a magnetar with persistent emission as pointed out by \citet{gav08}. In the
quiescent state, our flux measurement indicates a pulsar luminosity
$L_{X,0.5-10}^{\rm PSR}=2.6\times10^{34}$\,ergs\,s$^{-1}$. As mentioned in the
introduction, the original 19\,kpc distance estimate implies enormous X-ray
efficiencies for the pulsar and the PWN. The new distance estimate of 6\,kpc
greatly reduces the pulsar luminosity by an order of magnitude. Now the pulsar
has an X-ray efficiency $\eta_{\rm PSR}\equiv L_X^{\rm PSR}/\dot{E}\approx0.3\%$, 
which is no longer an outlier, but rather typical among young neutron stars
\citep[see][]{kp08}. For the PWN, the luminosity $L_{X,0.5-10}^{\rm PWN}=1.4
\times10^{35}$\,ergs\,s$^{-1}$ and efficiency $\eta_{\rm PWN}\approx2\%$ are
still larger than many others, but not exceptional. Also,
$\eta_{\rm PWN}/\eta_{\rm PSR}=6$ is close to the mean value of 4 obtained by
\citet{kp08}. Our results indicate that the X-ray properties of PSR
J1846--0258 in quiescence are not much different from an ordinary young neutron
star's. As compared to the Crab pulsar, PSR J1846--0258 has a higher surface
magnetic field but its nebular field is much weaker. This seems to imply no
direct correlation between the magnetic field strengths on a neutron star
surface and in its surrounding PWN.

Although the exact nature of the arc-like feature remains unknown, in either the
equatorial torus or equatorial wisp interpretation, the pulsar spin orientation 
can still be inferred from the spatial modeling. \citet{liv06} propose a
scenario in which the angle between the spin and magnetic axis
is $\alpha\approx9\arcdeg$ in order to explain the pulsar's braking index in
terms of the vacuum-dipole model \citep{mel97}. If this is real, then the
spin inclination $\zeta=63\arcdeg$ obtained from the torus fitting indicates
a magnetic inclination $\beta=\zeta-\alpha\approx 54\arcdeg$. The large $\beta$
is consistent with the non-detection of radio pulsations. This picture also
predicts that no pulsed $\gamma$-rays will be observable according to the
\citet{ry95} model. Future GLAST observations will address this.

\section{CONCLUSIONS}
In this study, we have carried out detailed spatial and spectral
analysis of the PWN in Kes~75. The deep Chandra observations reveal the
axisymmetric structure of the PWN with jet and torus features. We
deduce an average nebula field $B\sim15\mu$\,G from the X-ray emissivity
and interpret the arc-like structure in the northeast as equatorial outflow,
either a toroidal termination shock or a wisp. We further apply a torus
fitting scheme and show that the arc has a radius of $10.1\arcsec\pm0.4\arcsec$,
inclined at $62\pm2\arcdeg$ to the plane of the sky with the symmetry axis
at a position angle $207\arcdeg \pm8 \arcdeg$. The spatially resolved spectrum of
the PWN shows that regions along the jet direction have systematically
harder spectrum than the surroundings, suggesting a high velocity in the polar
outflow. We also deduce a minimum flow velocity of 0.4$c$ based on the
non-detection of a counter jet. There are two bright clumps aligned with
the jet, which we propose due to bow shocks of the polar outflow in the
supernova ejecta. Over a period of 6 years, the northern clump and the
inner jet demonstrate strong variability, which could be attributed to
MHD instabilities. Accompanying the recent outburst, the pulsar luminosity has
increased by 6 times and shows substantial softening in the spectrum. We
also find an emerging thermal component in the pulsar spectrum,
which could be related to the thermal afterglow of the bursts. The 2000
observation suggests a pulsar X-ray efficiency of 0.3\%. This value is not
particularly large among young neutron stars. In this aspect, PSR J1846--0258
was a rather typical pulsar in X-rays during quiescence.

Further investigation of the PWN properties will require high resolution
multi-wavelength observations. Also, long-term monitoring of the source would be
fruitful. The evolution of the thermal component in the pulsar spectrum can
indicate any physical changes of the hotspot on the neutron star surface, 
thus probing the nature of the bursts. Moreover, as this is the only known example
of magnetar-like bursts inside a PWN, any flux variation in the nebula
that is associated with the bursts could reveal the energy transport in the
system. This connection provides a unique opportunity to improve our understanding
of both PWN and magnetar physics.

\acknowledgements
We thank Mallory Roberts for useful discussions.
P.O.S. acknowledges support from NASA contract NAS8-39073 and Chandra
Grant G06-7053X. B.M.G. acknowledges the support of an Australian Research
Council Federation Fellowship.

{\it Facilities:} \facility{CXO (ACIS)}

\clearpage

\clearpage

\begin{deluxetable}{lr}
\tablecaption{BEST-FIT TORUS PARAMETERS.\label{tab1}}
\tablewidth{200pt}
\tabletypesize{\scriptsize}
\tablehead{\colhead{Parameter} & \colhead{Value}}
\startdata
$\Psi$ & $207\arcdeg\pm8\arcdeg$ \\
$\zeta$ & $62\arcdeg\pm5\arcdeg$ \\
$R$ & $10\arcsec\pm1\arcsec$ \\
blur  & 1\arcsec\ (fixed) \\
$\beta$ & $0.76\pm0.05$ \\
Point Source & 61.6$\times 10^3$ cts \\
Torus & 26.2$\times 10^3$ cts \\
Compact Nebula & 9.1$\times 10^3$ cts \\
Northern Jet & 12.2$\times 10^3$ cts \\ 
Southern Jet & 6.4$\times 10^3$ cts  
\enddata
\end{deluxetable}

\begin{deluxetable}{l|cc|cc||cc|cc}
\tablecaption {SPECTRAL FITS TO THE PWN.\label{tab2}}
\tablewidth{0pt}
\tabletypesize{\scriptsize}
\tablehead{\colhead{} &
\colhead{$\Gamma$} & \colhead{$f_{0.5-10}^{\rm abs}$} \vline &
\colhead{$\Gamma$} & \colhead{$f_{0.5-10}^{\rm abs}$} \vline\vline &
\colhead{$\Gamma$} & \colhead{$f_{0.5-10}^{\rm abs}$} \vline &
\colhead{$\Gamma$} & \colhead{$f_{0.5-10}^{\rm abs}$}}
\startdata
& \multicolumn{4}{c}{2000} \vline & \multicolumn{4}{c}{2006} \\
$N_{\rm H}$ & \multicolumn{2}{c}{$3.9\pm0.1$} & \multicolumn{2}{c}{4.0 (fixed)}
& \multicolumn{2}{c}{$4.06\pm0.07$} &  \multicolumn{2}{c}{4.0 (fixed)} \\
region 1& $2.3\pm0.3$ & $2.1\pm0.2$ & $2.3\pm0.3$ & $2.1\pm0.2$ & $2.10\pm0.13$ & $2.7\pm0.1$ & $2.07\pm0.12$ & $2.7\pm0.1$ \\
region 2& $2.1\pm0.3$ & $2.9\pm0.2$ & $2.2\pm0.2$ & $2.9\pm0.2$ & $2.05\pm0.11$ & $3.5\pm0.1$ & $2.02\pm0.10$ & $3.5\pm0.1$ \\
region 3& $2.0\pm0.1$ & $12.9\pm0.5$ & $2.1\pm0.1$ & $12.8\pm0.5$ & $2.06\pm0.06$ & $13.6\pm0.2$ & $2.04\pm0.05$ & $13.7\pm0.2$ \\
region 4& $1.7\pm0.2$ & $3.4\pm0.3$ & $1.8\pm0.2$ & $3.4\pm0.3$ & $2.09\pm0.11$ & $3.5\pm0.1$ & $2.07\pm0.10$ & $3.5\pm0.1$ \\
region 5& $1.9\pm0.1$ & $9.9\pm0.4$ & $2.0\pm0.1$ & $9.9\pm0.4$ & $2.01\pm0.07$ & $9.4\pm0.2$ & $1.99\pm0.06$ & $9.4\pm0.2$ \\
region 6& $1.8\pm0.1$ & $19.5\pm0.6$ & $1.8\pm0.1$ & $19.4\pm0.6$ & $1.75\pm0.05$ & $21.1\pm0.3$ & $1.73\pm0.04$ & $21.2\pm0.3$ \\
region 7& $1.9\pm0.1$ & $16.7\pm0.5$ & $2.0\pm0.1$ & $16.6\pm0.5$ & $1.94\pm0.05$ & $15.9\pm0.3$ & $1.91\pm0.04$ & $16.0\pm0.3$ \\
region 8& $1.7\pm0.2$ & $4.6\pm0.3$ & $1.7\pm0.2$ & $4.6\pm0.3$ & $1.99\pm0.10$ & $4.3\pm0.1$ & $1.97\pm0.09$ & $4.3\pm0.1$ \\
region 9& $2.4\pm0.3$ & $1.5\pm0.2$ & $2.4\pm0.3$ & $1.5\pm0.2$ & $2.08\pm0.15$ & $2.1\pm0.1$ & $2.05\pm0.14$ & $2.1\pm0.1$ \\
region 10& $1.8\pm0.2$ & $6.7\pm0.4$ & $1.8\pm0.2$ & $6.6\pm0.4$ & $1.99\pm0.08$ & $6.0\pm0.2$ & $1.96\pm0.08$ & $6.0\pm0.2$ \\
region 11& $1.8\pm0.1$ & $12.6\pm0.5$ & $1.8\pm0.1$ & $12.5\pm0.5$ & $1.83\pm0.06$ & $12.3\pm0.2$ & $1.80\pm0.05$ & $12.4\pm0.2$ \\
region 12& $1.9\pm0.1$ & $10.8\pm0.4$ & $1.9\pm0.1$ & $10.7\pm0.4$ & $1.91\pm0.06$ & $12.0\pm0.2$ & $1.88\pm0.05$ & $12.1\pm0.2$ \\
region 13& $2.0\pm0.2$ & $3.7\pm0.3$ & $2.0\pm0.2$ & $3.7\pm0.3$ & $2.10\pm0.10$ & $4.0\pm0.1$ & $2.07\pm0.09$ & $4.0\pm0.1$ \\
region 14& $1.8\pm0.2$ & $3.9\pm0.3$ & $1.9\pm0.2$ & $3.8\pm0.3$ & $1.83\pm0.09$ & $5.6\pm0.2$ & $1.81\pm0.08$ & $5.6\pm0.2$ \\
region 15& $1.7\pm0.1$ & $8.4\pm0.4$ & $1.7\pm0.1$ & $8.3\pm0.4$ & $1.86\pm0.07$ & $8.8\pm0.2$ & $1.84\pm0.06$ & $8.8\pm0.2$ \\
region 16& $2.1\pm0.2$ & $2.8\pm0.2$ & $2.1\pm0.2$ & $2.8\pm0.2$ & $2.20\pm0.12$ & $2.8\pm0.1$ & $2.17\pm0.12$ & $2.8\pm0.1$ \\
region 17& $1.7\pm0.2$ & $7.1\pm0.4$ & $1.7\pm0.2$ & $7.1\pm0.4$ & $1.93\pm0.08$ & $6.7\pm0.2$ & $1.90\pm0.07$ & $6.7\pm0.2$ \\
region 18& $1.4\pm0.2$ & $5.6\pm0.4$ & $1.4\pm0.2$ & $5.6\pm0.4$ & $1.57\pm0.10$ & $5.0\pm0.2$ & $1.54\pm0.09$ & $5.0\pm0.2$ \\
region 19& $1.7\pm0.2$ & $5.2\pm0.3$ & $1.7\pm0.2$ & $5.2\pm0.3$ & $1.65\pm0.09$ & $5.6\pm0.2$ & $1.63\pm0.08$ & $5.7\pm0.2$ \\
region 20& $1.9\pm0.2$ & $4.7\pm0.3$ & $1.9\pm0.2$ & $4.7\pm0.3$ & $1.99\pm0.09$ & $5.2\pm0.2$ & $1.96\pm0.08$ & $5.2\pm0.2$ \\
total & $1.84\pm0.07$ & $141\pm2$ & $1.88\pm0.03$ & $139\pm2$ & $1.93\pm0.03$ & $143\pm0.8$ & $1.89\pm0.02$ & $144\pm0.8$
\enddata
\tablecomments{$N_{\rm H}$ is in the unit of $10^{22}\mathrm{cm}^{-2}$ and $f_{0.5-10}^{\rm abs}$ is the absorbed flux in the 0.5-10\,keV range in the unit of
$10^{-13}\,\mathrm{ergs\,s^{-1}\,cm^{-2}}$.
The uncertainties quoted are 90\% confidence intervals.}
\end{deluxetable}

\begin{deluxetable}{c|cccccccc}
\tablecaption {SPECTRAL FITS TO PSR J1846--0258\label{tab3}}
\tablewidth{0pt}
\tabletypesize{\scriptsize}
\tablehead{\colhead{Model} & \colhead{$N_{\rm H}$} &
\colhead{$\Gamma$} & \colhead{$f_{0.5-10}^{\rm abs,PL}$} & \colhead{$f_{0.5-10}^{\rm unabs,PL}$} &
\colhead{$k$T (keV)} & \colhead{$f_{0.5-10}^{\rm abs,BB}$} & \colhead{$f_{0.5-10}^{\rm unabs,BB}$} & \colhead{$\chi^2/\nu$}}
\startdata
\sidehead{2000} \\
PL & $4.0^\dagger$ & $1.1\pm0.1$ & $0.42\pm0.02$ & $0.61\pm0.03$ &
\nodata & \nodata & \nodata & 36.4/40 \\
PL+BB & $4.0^\dagger$ & $1.0^{+0.8}_{-0.3}$ & $0.42\pm0.02$ & $0.61\pm0.04$ &
$0.4\pm0.4$ & $0.001^{+0.003}_{-0.001}$ & $0.01^{+0.03}_{-0.01}$ & 35.6/38 \\
\sidehead{2006} \\
PL & $4.0^\dagger$ & $1.86\pm0.02$ & $1.64\pm0.03$ & $3.7\pm0.1$ &
\nodata & \nodata & \nodata & 148.8/136 \\
PL+BB & $4.0^\dagger$ & $1.9\pm0.1$ & $1.3\pm0.3$ & $3.1\pm0.6$ &
$0.9\pm0.2$ & $0.17\pm0.02$ & $0.32\pm0.04$ & 132.1/134
\enddata
\tablenotetext{\dag}{ -- held fixed in the fit.} 
\tablecomments{$N_{\rm H}$ is in the unit of $10^{22}\mathrm{cm}^{-2}$.
$f_{0.5-10}^{\rm abs,PL}$ and $f_{0.5-10}^{\rm unabs,PL}$ are the absorbed and unabsorbed
fluxes respectively for the powerlaw component, while $f_{0.5-10}^{\rm abs,BB}$
and $f_{0.5-10}^{\rm unabs,BB}$ are the absorbed and unabsorbed fluxes for the blackbody
component. All the reported fluxes are in units of
$10^{-11}\,\mathrm{ergs\,s^{-1}\,cm^{-2}}$ in the 0.5-10\,keV range.
The uncertainties quoted are 90\% confidence intervals.}
\end{deluxetable}

\begin{deluxetable}{l|l|ccccc}
\tablecaption {SPECTRAL FITS TO THE JET AND CLUMP FEATURES\label{tab4}}
\tablewidth{0pt}
\tabletypesize{\scriptsize}
\tablehead{
\colhead{Region} & \colhead{Model} & \colhead{$N_{\rm H}$} &
\colhead{$\Gamma$/ $k$T (keV)} & \colhead{$f_{0.5-10}^{\rm abs}$}
& \colhead{$f_{0.5-10}^{\rm unabs}$} & \colhead{$\chi^2/\nu$} }
\startdata
\sidehead{2000}
northern clump & PL & $4.0^\dagger$ & $1.6\pm0.1$ & $15.0\pm0.7$ & $28.2\pm1.4$ &14/16 \\
inner jet & PL & $4.0^\dagger$ & $1.6\pm0.5$ & $1.2\pm0.2$ & $2.2\pm0.4$ & 6.3/9 \\
outer jet & PL & $4.0^\dagger$ & $1.1\pm0.3$ & $4.0\pm0.4$ & $6.0\pm0.6$ & 5.5/9 \\
southern clump & PL & $4.0^\dagger$ & $1.7\pm0.3$ & $2.9\pm0.3$ & $5.9\pm0.7$ & 2.9/9 \\
\sidehead{2006}
northern clump & PL & $4.0^\dagger$ & $1.6\pm0.07$ & $15.2\pm0.4$ & $28.7\pm0.7$ & 109/161 \\
 & PL & $3.8\pm0.3$ & $1.5\pm0.14$ & $15.5\pm0.4$ & $27.4\pm0.7$ & 108/160 \\
 & Bremss. & $4.0^\dagger$ & $13^{+4}_{-2}$ & $14.5\pm0.4$ & $26.5\pm0.7$ & 112/161 \\
northern clump 1 & PL & $4.0^\dagger$ & $1.5\pm0.1$ & $6.6\pm0.3$ & $11.6\pm0.5$ & 42/56 \\
northern clump 2 & PL & $4.0^\dagger$ & $1.6\pm0.1$ & $8.8\pm0.3$ & $17.1\pm0.6$ & 43/56 \\
inner jet & PL & $4.0^\dagger$ & $1.7\pm0.2$ & $2.1\pm0.1$ & $4.3\pm0.2$ & 25/56 \\
 & PL & $4.6^{+1.0}_{-0.8}$ & $2.0\pm0.4$ & $2.0\pm0.1$ & $5.2\pm0.3$ & 24/56 \\
outer jet & PL & $4.0^\dagger$ & $1.3\pm0.2$ & $3.4\pm0.2$ & $5.4\pm0.3$ & 28/56 \\
 & PL & $4.8^{+0.9}_{-0.8}$ & $1.6^{+0.4}_{-0.3}$ & $3.2\pm0.2$ & $6.3\pm0.4$ & 26/55 \\
southern clump & PL & $4.0^\dagger$ & $1.5\pm0.2$ & $2.7\pm0.2$ & $4.8\pm0.3$ & 38/56 \\
 & PL & $3.4^{+1.0}_{-0.8}$ & $1.3\pm0.4$ & $2.8\pm0.2$ & $4.4\pm0.3$ & 37/55 \\
 & Bremss. & $4.0^\dagger$ & $18^{+20}_{-9}$ & $2.6\pm0.2$ & $4.6\pm0.3$ & 39/56
\enddata
\tablenotetext{\dag}{ -- held fixed in the fit.}
\tablecomments{$N_\mathrm{H}$ is in the unit of $10^{22}$\,cm$^{-2}$, the absorbed 
($f_{0.5-10}^{\rm abs}$) and unabsorbed fluxes ($f_{0.5-10}^{\rm unabs}$) are in 
units of $10^{-13}\,\mathrm{ergs\,s^{-1}\,cm^{-2}}$ in the 0.5-10\,keV range.
The uncertainties quoted are 90\% confidence intervals.}
\end{deluxetable}

\end{document}